\documentstyle[amssymb,aps,12pt]{revtex}

\begin{document}
\author{M. S. Hussein and M. P. Pato}
\address{Nuclear Theory and Elementary Particle Phenomenology Group\\
Depto. de F\'{i}sica Nuclear, Instituto de Fisica, Universidade de S\~{a}o\\
Paulo\\
C.P. 66318, 05315-970 S\~{a}o Paulo, S.P., Brazil}
\title{Matrix Elements of Random Operators and Discrete Symmetry Breaking in Nuclei%
\thanks{%
Supported in part by the CNPq and FAPESP (Brazil)}}
\maketitle

\begin{abstract}
It is shown that several effects are responsible for deviations of the
intensity distributions from the Porter-Thomas law. Among these are genuine
symmetry breaking, such as isospin; the nature of the transition operator;
truncation of the Hilbert space in shell model calculations and missing
transitions.
\end{abstract}

Random matrix theory (RMT)\cite{Meht,Brody,Weide} has had a wide application
in the description of the statistical properties of eigenvalues and
eigenfunctions of complex many-body systems. More general ensembles have
also been considered\cite{Dyson}, in order to cover situations that depart
from the universality classes of RMT. One such class of ensembles is the
so-called deformed Gaussian orthogonal ensemble (DGOE)\cite
{Pato1,Pato2,Pato3}. These ``intermediate '' ensembles are particularly
useful when one wants to study the breaking of a given symmetry in a
many-body system such as the atomic nucleus. When discrete symmetries such
as isospin, parity and time reversal are violated in a complex many-body
environment, one relies on a description based on the transition between one
GOE into two coupled GOE 's in the first two cases\cite{Pato3,Guhr} and a
GOE into a Gaussian Unitary Ensemble (GUE) in the last case.

An important test of the statistical behavior of a complex many-body system
is the distribution of the eigenvectors components. If fully chaotic, the
system obeys the Porter-Thomas law\cite{Porter}. This is verified in a
variety of cases and using shell-model calculation\cite{Brown,Dias,Zelev}.
Some deviations from this distribution were reported which may reflect
several of many effects: the limited applicability of the ergodic theorem,
some missing strengths and otherwise unknown broken hidden symmetry.

A very important test of the RMT in the context of discrete symmetry
breaking has been supplied by Mitchell and collaborators at TUNL where
isospin mixing was investigated in the spectrum of $^{26}$Al\cite{Mitch0}.
Here, full experimental knowledge of the discrete spectrum up to 8 Mev of
excitation energy is almost complete ( about 100 states). Further, owing to
the odd-odd character of this isotope (13 protons+13 neutrons) one expects
to find both T=0 and T=1 states even at very low excitation energies.
Therefore, $^{26}$Al is a very convenient laboratory to study isospin
breaking in the nuclear states. Only recently, the study of the transition
distribution was made following some suggestions concerning the nature of
the manifestation of a discrete symmetry breakdown in the distribution made
by Alhassid and Levine\cite{Alha} and Hussein and Pato\cite{Pato1}. The
recent work of TUNL showed a great deviation of the distribution from the
pure Porter-Thomas form\cite{Mitch}. A possible cause of this may be a
collective state responsible for the T-mixing. No clear explanation, however
is given. One reason for this is the lack of a detailed theoretical study of
such deviations. Here we supply a model that allows a better understanding
of the effect of symmetry breaking on the transition distribution.

In the following we show that it is possible to introduce a random operator
whose matrix elements simulate, in some way, the behavior of observables of
complex physical systems. In the construction of this operator, we will be
guided by the idea that when a system undergoes a chaos-order transition, a
quantity that has a key role in determining the statistical behavior of the
matrix elements of an operator, is the expectation value of its commutator
with the Hamiltonian. This is implied by the equation

\begin{equation}
\left( E_{l}-E_{k}\right) <E_{l}\mid O\mid E_{k}>=<E_{k}\mid [O,H]\mid
E_{k}>=i\hbar \frac{d}{dt}<E_{l}\mid O\mid E_{k}>  \label{eq 1}
\end{equation}
where the last equality was obtained using Schr\"{o}dinger equation and
assuming an operator with no explicit dependence on time. Eq.$\left( \ref{eq
1}\right) $ clearly shows that the commutator supplies the connection
between the matrix elements of an observable and its behavior as a function
of time.

To see how this is reflected in the statistical distribution, we consider an
observable $O$ which becomes a conserved quantity in the regular regime. The
distribution of the matrix elements of $O$ will undergo a transition from
the Porter-Thomas law, in the chaotic regime, to a distribution
corresponding to the singular $<E_{l}\mid O\mid E_{k}>$ $\varpropto \delta
_{kl},$ in the regular regime, since the last term in $\left( \ref{eq 1}%
\right) $ is zero in this latter case. When considering the question of
violation of a discrete symmetry in a complex system one resorts to a
description involving the trannsition 2GOE$\rightarrow 1$GOE, where it is
assumed that in the first limit (2GOE) one has a conserved quantum number
that labels two sets of levels (e.g., parity or T=0 and T=1 states). The
opposite limit (1GOE) involves the complete breakdown of the symmetry. Then,
in general, for some weak symmetry violation, one has two coupled GOE's. The
operator $O$ may have non-zero matrix element for the transitions belonging
to the two different GOE blocks of the Hamiltonian matrix. This causes a
deviation of the distribution from the P-T law. However, even if the
symmetry is not broken, pieces of the operator $O$ may also couple these
states and will also result in deviation from the P-T law. Here is the
dilemma which we will address in the following: how to distinguish genuine
symmetry breaking effects from false ones related to the nature of $O$
through the study of the deviations from the P-T law. Of course, missing
transitions (not counted in the analysis) also result in a significant
deviation.

To define the ensembles of random matrices we are going to work with, we
follow the construction based on the Maximum Entropy Principle\cite{Pato1},
that leads to a random Hamiltonian which can be cast into the form

\begin{equation}
H=H_{0}+\lambda H_{1},  \label{eq 1b}
\end{equation}
where $\lambda $ ($0\leq \lambda \leq 1)$ is the parameter that controls the
chaoticity of the ensemble. For $\lambda =1,$ $H=H^{GOE},$ the fully chaotic
situation, whereas for $\lambda =0,$ we have a different situation defined
by $H_{0}$ (e.g. two uncoupled GOE's). Since we are specifically interested
in the transition from GOE to two uncoupled GOE's, we write\cite{Pato1,Pato2}

\begin{equation}
H_{0}=PH^{GOE}P+QH^{GOE}Q  \label{eq 4}
\end{equation}
and

\begin{equation}
H_{1}=PH^{GOE}Q+QH^{GOE}P  \label{eq 5}
\end{equation}
where $P=\sum\limits_{i=1}^{M}P_{i},$ $Q=1-P$ and $P_{i}=\mid i><i\mid ,$ $%
i=1,\ldots ,N$ are projection operators. Therefore $H_{0}$ is a two blocks
diagonal matrix of dimensions $M$ and $N-M$ and each block is by
construction a GOE random matrix$.$ It is easily verified that $H=H^{GOE}$
for $\lambda =1.$

There are two main classes of operators to be considered. The first class is
the set of symmetry conserving, $O_{C}$ operators which in the limit of
conserved symmetry, $\lambda =0$, i.e., two GOE's in our model, have
elements only inside the blocks while, the other class is that of the
symmetry breaking $O_{B}$ ones which in the same limit have elements only
between the blocks. Starting with the $O_{C}$ ones, they can be simulated in
the model by an operator of the form

\begin{equation}
O_{C}=\sum_{i=0}^{N}P_{i}H^{GOE}P_{i}.  \label{eq 6}
\end{equation}
To see how it acts, it is convenient to separate the sum in Eq. $\left( \ref
{eq 6}\right) $ into two parts in which the first $M$ terms define the
operator $O_{CP}$ and the other $N-M$ define the operator $O_{CQ}$ , by
construction $O_{C}=O_{CP}+O_{CQ}.$ The commutator with the Hamiltonian can
then be written as

\begin{equation}
\lbrack O_{CP},PHP]+[O_{CQ},QHQ]+\lambda \left(
O_{CP}PHQ+O_{CQ}QHP-QHPO_{CP}-PHQO_{CQ}\right) .  \label{eq 10}
\end{equation}
The first and the second terms are block diagonal while those multiplied by $%
\lambda $ are block anti-diagonal. In the limit $\lambda \rightarrow 0,$ the
eigenstates become localized in the blocks and only the transitions between
states inside the blocks $PHP$ and $QHQ$ survive.

In the case of symmetry breaking operators $O_{B}$, they can be simulated
simply by $H_{1},$ i.e., $O_{B}=H_{1}$. In fact, its commutator with the
Hamiltonian is

\begin{equation}
\left[ H,O_{B}\right] =PHPHQ+QHQHP-PHQHQ-QHPHP  \label{eq 14}
\end{equation}
which is block anti-diagonal. So, as consequence, when $\lambda \rightarrow
0 $ this operator induces transitions only between the blocks$.$

The above properties of these operators make them very convenient to study
the situation when a selection rule becomes operative in a transition
towards two coupled GOE's, a scenario appropriate to investigate discrete
symmetry violation, e.g., isospin, in nuclei. In order to make our model
more flexible we are going to study distributions of matrix elements of the
generic operator

\begin{equation}
O=\left( 1-q\right) O_{C}+qO_{B}  \label{eq 18}
\end{equation}
where $q$ varies between $0$ and $1$. With this more generic form the model
depends on two parameters, the parameter $\lambda $ of the Hamiltonian which
may be fixed by the fitting the eigenvalues distribution and the parameter $%
q $ that selects the operator.

Following the standard procedure, we first construct, with the operator $O,$
the normalized vector

\begin{equation}
\mid \alpha _{k}>=O\mid E_{k}>  \label{eq 22}
\end{equation}
where $\mid E_{k}>$ with $k=1,\ldots ,N$ is an eigenvector of the
Hamiltonian. From these $N$ vectors we define the matrix elements

\begin{equation}
T_{kl}=<E_{l}\mid \alpha _{k}>  \label{eq 26}
\end{equation}
which are the quantities to be statistically analyzed. It is convenient to
work with $\mid T_{kl}\mid ^{2},$ and perform a local average that extracts
secular variation with the energies. Thus we introduce the quantities

\begin{equation}
y_{kl}=\frac{\mid T_{kl}\mid ^{2}}{\left\langle \mid T_{kl}\mid
^{2}\right\rangle }  \label{eq 30}
\end{equation}
where the average is done by using a Gaussian filter of variance equal to 2
and, as it has become standard in the analysis of these quantities, we
histogram their logarithm.

In Fig. 1, it is shown the numerical simulations performed with the symmetry
conserving operator, $O_{C}$, for four values of the parameter $\lambda .$
The striking characteristic of these plots is the strange dependence of
these statistics on the chaoticity parameter. The evolution of the
distribution with $\lambda $ is clearly seen to be P-T$\rightarrow $ non P-T$%
\rightarrow $ P-T. Thus, great care must be taken when confronting deviation
from P-T from genuine symmetry breaking. Very strong symmetry breaking could
lead to a P-T distribution.

In Fig. 2, the results are shown for the distribution of the matrix element
of the general operator $O$. The chaoticity parameter was fixed at value $%
\lambda =0.032$ which is so fixed to account for the isospin breaking seen
in the eigenvalue distribution of Ref. \cite{Mitch0}, and the parameter $q,$
that measures the ``symmetry deformation'' of $O$, is varied. At the above
value of $\lambda $, we expect to be near the case in which the two GOE's
are nearly decoupled. We see from the figure that the distributions are
strongly dependent on the parameter $q$. Thus, very weak symmetry violation
in the system may result in a strong deviation of the transition strength
distribution from a P-T one owing to the nature of the transition operator.

One important question from the experimental side is the effect of missing
strengths in the data analysis. The experimental apparatus may have a
minimum below which strengths are not detected or, else, it may not be able
to resolve a relatively weak strength staying close to a strong one. The
fact that the data is not complete may result in a distortion of the final
distribution. This may be the case, for example, of the measured
distribution of Ref. \cite{Mitch}, in which besides the shift to the left,
in qualitative agreement with our theoretical distributions, the data
exhibits a peak higher than the Porter-Thomas distribution. In order to
check this explanation, we have performed a simulation of the effect in our
model, by eliminating, from the distribution, matrix elements below some
threshold value. The result obtained is shown as the histogram in Fig 3
together with the P-T distribution. Indeed, it is seen in the distorted
distribution the same qualitative trend as the data analysis of Ref. \cite
{Mitch}, namely, an enhancement of the maximum which is shifted to the left.
Therefore, missing transitions {\it also }lead to significant deviation from
the P-T law.

It is interesting to observe at this point that all the deviations discussed
above can be represented most adequately by the sum of two $\chi ^{2}$%
-distributions as was suggested earlier\cite{Pato2}.

Before ending, it is of interest to briefly discuss some recent result of
extensive shell model calculation of the intensity distribution. Hamoudi,
Nazmitdinov and Alhassid\cite{Hamou} made statistical fluctuation analysis
of electromagnetic transition intensities in A=60 nuclei using the shell
model code OXBASH. They specifically looked at E2 and M1 transitions among
T=0,1 states. They found that B(E2) distributions are well described by the
P-T law, independently of the value of T$_{z}=\frac{N-Z}{2}$ . However, they
point out that B(M1) distributions depend strongly on T$_{z}$: T$_{z}=1$
nuclei obey the P-T law while T$_{z}=0$ show significant deviation from the
P-T prediction.

In light of our discussion and considerations above, the fact that very
small violation of T is expected we conjecture that the deviation in Ref. 
\cite{Hamou} is a ``false '' symmetry breaking violation effect that arises
from the truncation of the basis in the calculation. Further study is
certainly needed to better understand the phenomenon.

In conclusion, we have investigated in this letter possible causes for the
strength intensity distributions to deviate from the Porter-Thomas law. We
can trace the deviations to: genuine discrete symmetry breaking, such as
isospin; the nature of the transition operator; truncation of the Hilbert
space (which requires dealing with effective transition operators that may
contain ``false '' symmetry violation terms) and ``missing transitions''. An
analysis of the data of Ref. \cite{Mitch} using our model is underway.

\section{Acknowledgment}

We thank Gary Mitchell for very useful suggestions and remarks.

\strut

\ {\bf Figure Captions:}

Fig. 1 Four histograms of the logarithm of the matrix elements distributions
of the random operator $O_{C}$ for the transition GOE$\rightarrow $2GOE's,
for the indicated values of the parameter $\lambda $. The calculations were
done with matrices of dimension $N=150$ and block sizes $M_{1}=80$ and $%
M_{2}=70$. The dashed line corresponds to P-T, the dotted line to the fit
with one $\chi ^{2}$- distribution and the solid one to a sum of two $\chi
^{2}$- distributions. See text for details.

Fig. 2 Four histograms of the logarithm of the matrix elements distributions
of the generic random operator $O$ for the case of the transition GOE$%
\rightarrow $2GOE's, for the indicated values of the parameter $q$. The
calculations were done with matrices of dimension $N=150$ and block sizes $%
M_{1}=80$ and $M_{2}=70$. The lines are as in Fig. 1.

Fig. 3 The effect of the missing transitions in the P-T distribution. The
calculations were done with matrices of dimension $N=100$. See text for
details.

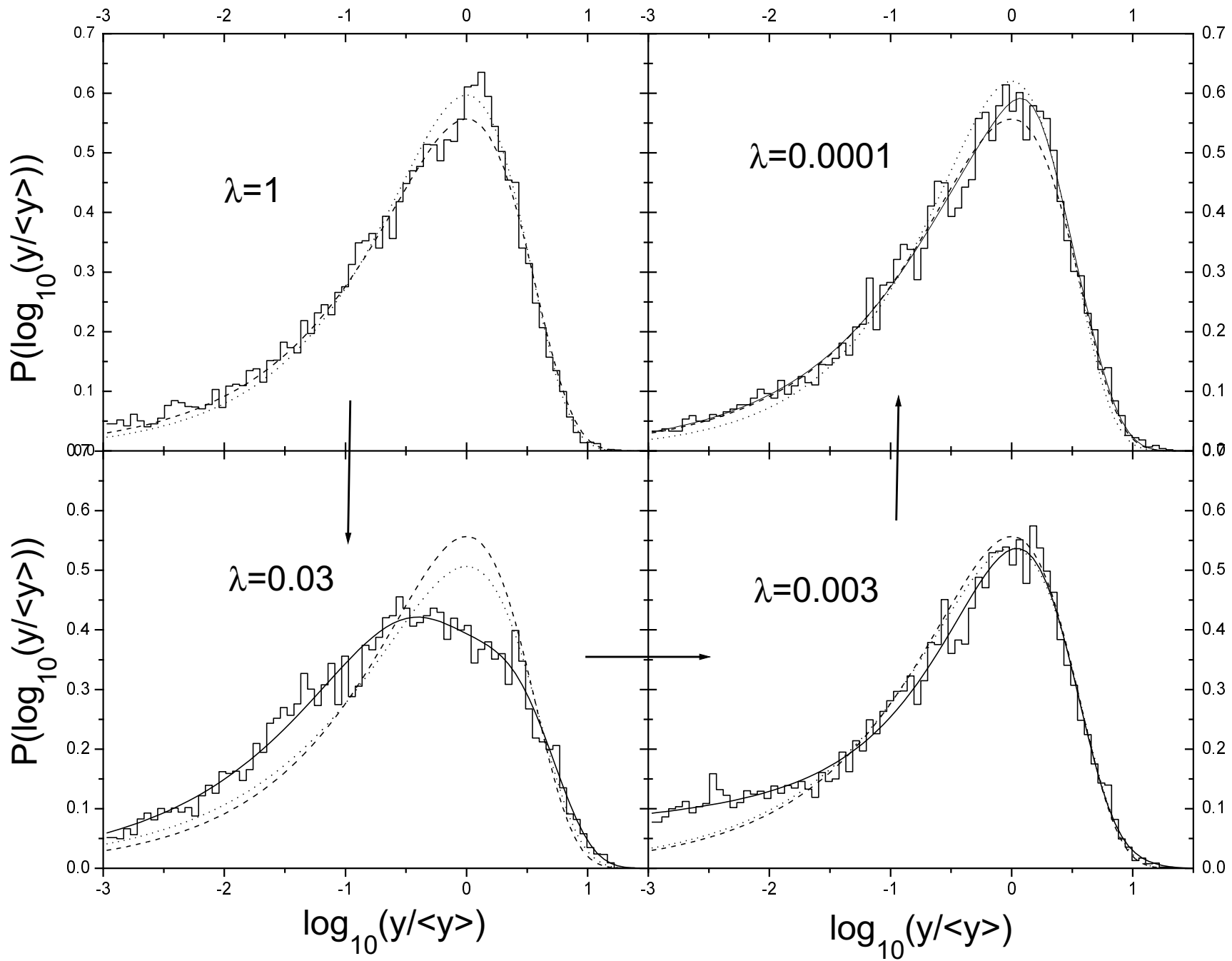
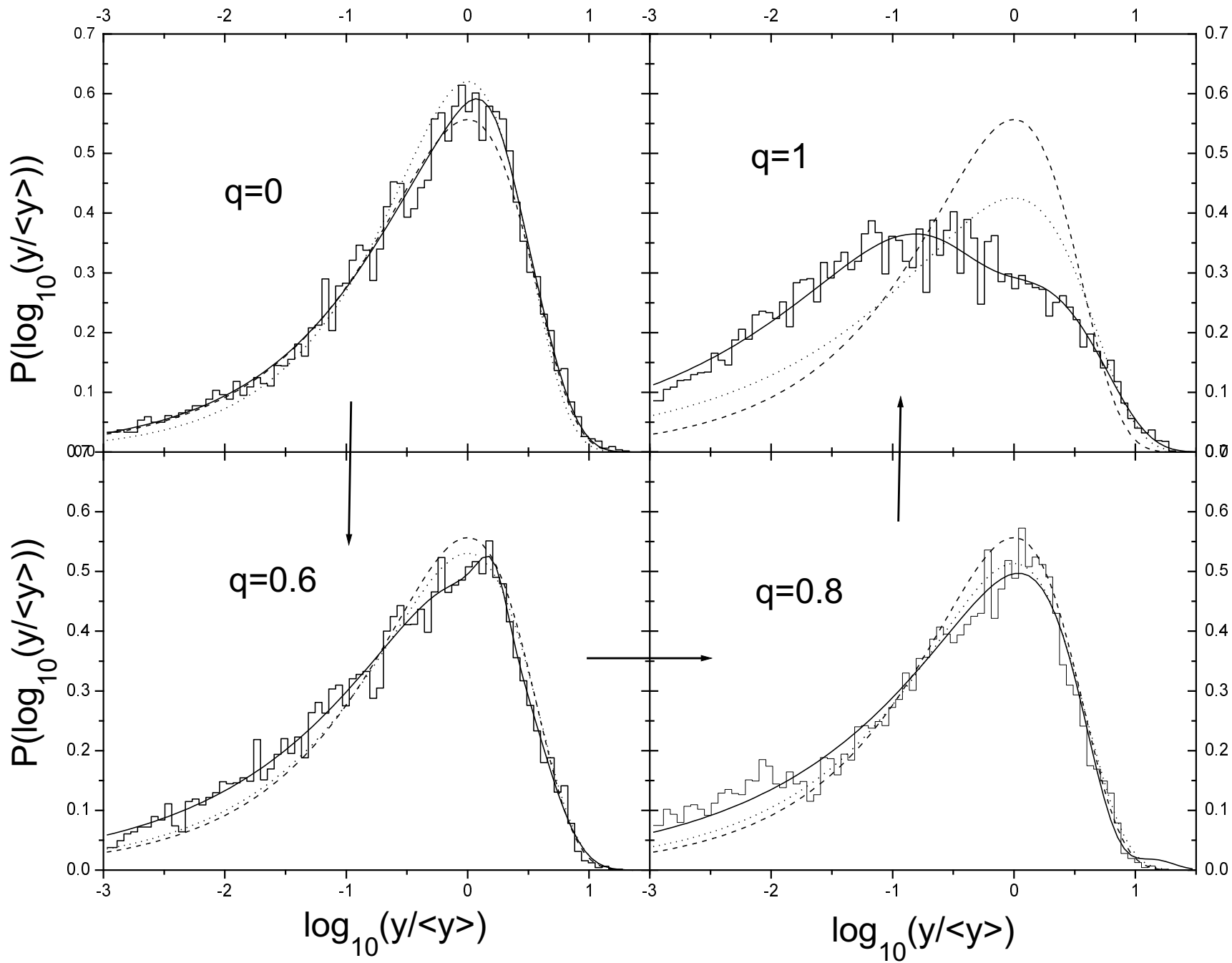
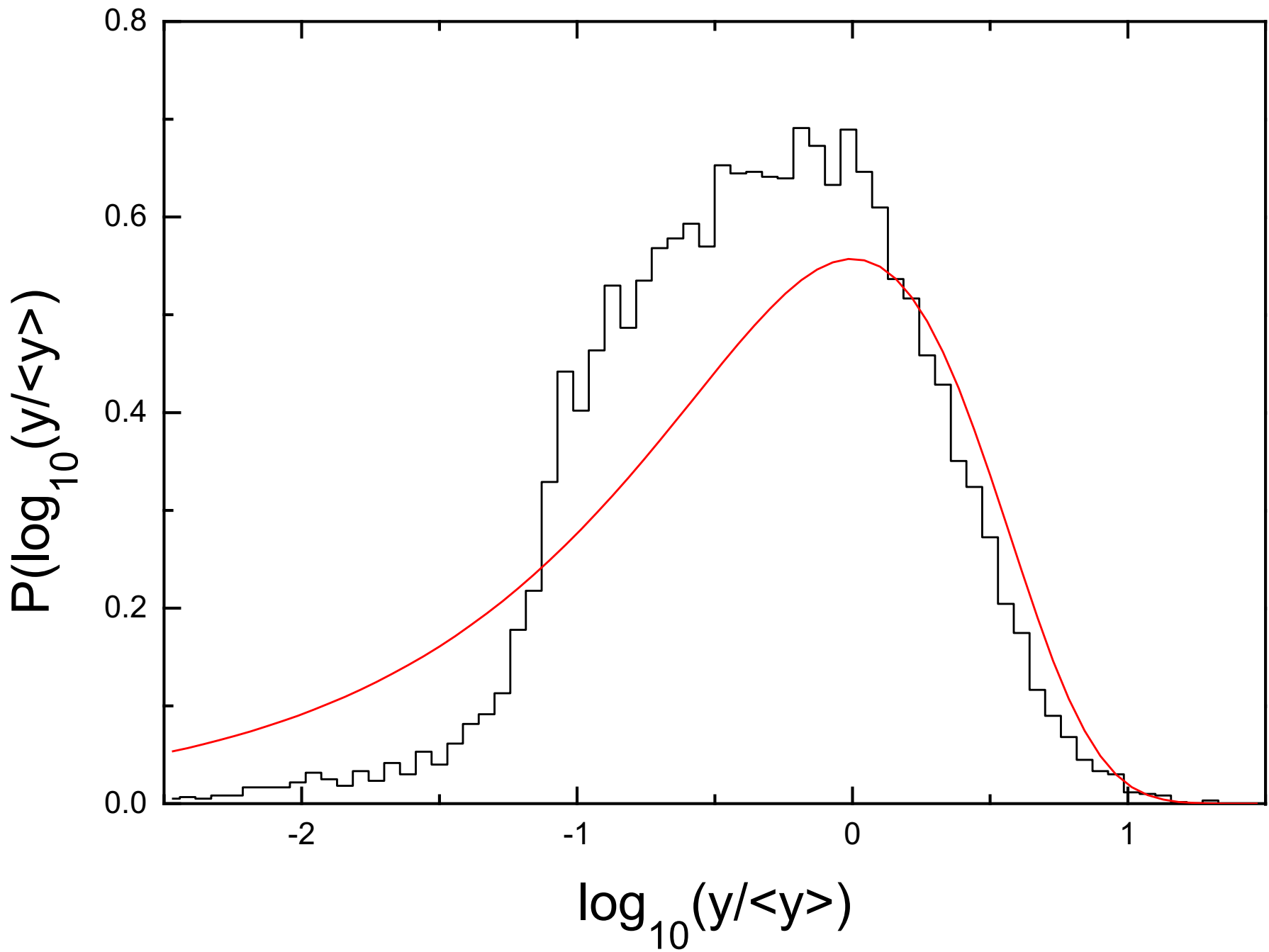
\end{document}